\documentclass{article}

\usepackage{arxiv}

\usepackage[utf8]{inputenc} 
\usepackage[T1]{fontenc}    
\usepackage{hyperref}       
\usepackage{url}            
\usepackage{booktabs}       
\usepackage{amsfonts}       
\usepackage{nicefrac}       
\usepackage{microtype}      
\usepackage{lipsum}
\usepackage{graphicx}
\usepackage[
style=apa,
]{biblatex}
\title{Biometric Technologies and the Law: Developing a Taxonomy for Guiding Policymakers}

\author{
 Luis Felipe M. Ramos \\
  School of Law\\
  University of Minho\\
  Braga, Portugal \\
  \texttt{lfelipe.sm@gmail.com} \\
}

\begin{document}
\maketitle
\begin{abstract}
Despite the increasing adoption of biometric technologies, their regulation has not kept up with the same pace, particularly with regard to safeguarding individuals' privacy and personal data. Policymakers may struggle to comprehend the technology behind biometric systems and their potential impact on fundamental rights, resulting in insufficient or inadequate legal regulation. This study seeks to bridge this gap by proposing a taxonomy of biometric technologies that can aid in their effective deployment and supervision. Through a literature review, the technical characteristics of biometric systems were identified and categorised. The resulting taxonomy can enhance the understanding of biometric technologies and facilitate the development of regulation that prioritises privacy and personal data protection.
\end{abstract}


\section{Introduction}
\label{sec:introduction}

Over the past few decades, there has been significant growth in the development and adoption of biometric technologies, which are utilised in various fields such as access control, border management, device security, and digital identification.

However, the growing concern over safeguarding individual privacy has become more prominent, particularly with advanced technologies like facial and emotion recognition. Since biometric data is inherently linked to an individual, any breach or exposure can lead to a violation of privacy and potentially cause long-lasting problems, as biometric data is irreplaceable.

Due to increased protection for personal data and the demand for comparable guarantees for international data transfers, many countries have started to legislate on the issue, providing their citizens with a minimum level of protection. This trend began after the adoption of the European Union's General Data Protection Regulation in 2018 (EU GDPR).

Most of these countries view privacy as a fundamental human right and have established a robust and all-encompassing legal framework that regulates this right based on their constitutional traditions. The Council of Europe's Convention No. 108 has played a crucial role in shaping this approach. Despite this, such extensive legislation on data privacy often falls short of achieving its intended goals \parencite{Reidenberg_2015}.

Nevertheless, most data protection legislation enacted so far makes little or no reference to more specific topics, such as biometric recognition. For example, the EU GDPR employs the word ‘biometric’ only six times throughout the entire regulation, all of which refer to ‘biometric data’. It is important to note that the EU GDPR, in its Article 9, does not provide a dedicated set of rules for biometric technologies but instead applies the same rules as those enacted for personal data revealing racial or ethnic origin, political opinions, religious or philosophical beliefs, or trade union membership, and genetic data, data concerning health or data concerning a natural person’s sex life or sexual orientation, all under the umbrella term of ‘special categories of personal data’.

The Brazilian general data protection law (Lei n. 13.709, de 14 de Agosto de 2018 - Lei Geral de Proteção de Dados) makes only one reference to ‘biometric data’ in Article 5º, II, when defining the concept of ‘sensitive personal data’. The California Consumer Privacy Act of 2018 (CCPA) (Cal. Civ. §1798.100), in its § 1798.140, presents five references to ‘biometric information’, all of them when defining that concept and the concept of ‘personal information’. Just like the EU GDPR, the CCPA includes ‘biometric information’ under the umbrella term of ‘sensitive personal information’, applying to it the same rules as the ones enacted for consumer’s social security, driver’s license, or passport number, their precise geolocation, their racial or ethnic origin, religious or philosophical beliefs, or union membership, genetic data, or their health, sex life or sexual orientation.

A significant aspect of the structure and language adopted in these legislations is that they induce biometric systems to be regulated by the same set of rules as the processing of personal data originating from other applications, ignoring the differences and complexities existing among these applications and the several types of biometric systems.

Another relevant aspect of this broad, almost generic, regulation is the consequential enactment of legislation banning or imposing moratoria or strict requirements on the use of some of these systems, as we have seen in some cities in the USA, such as San Francisco and Boston \parencite{Conger_Fausset_Kovaleski_2019}, and the EU \parencite{Heikkilä_2021}. This approach can also be considered problematic since it focuses on some issues of the technology (usually bias in the data collection process) and ignores its potential benefits if adequately regulated. And this regulation shall not only take into consideration social and ethical questions about its use, as different systems present different implications, requiring different legal responses, but also must be written in terms that every person impacted by the technology can understand it \parencite{Baecker_2022}.

Two non-exclusive aspects may cause the lack of adequate regulation of the matter: (i) the use of different expressions to designate the same things or the same expressions for different things; and (ii) difficulty by policymakers to understand the technology involved in the biometric systems and its potential consequences for the citizens' fundamental rights.

The first aspect has recently been addressed by an ISO and IEC Joint Technical Committee, which completed the harmonisation of the vocabulary of the subject field of biometrics through the revision and publication of the ISO/IEC 2382-37:2022 standard. This document intends to clarify the terms related to biometrics, providing a systematic description of the concepts used in that subject field.

The second aspect can be addressed by providing a new tool for understanding biometric systems by classifying some of their components. Approaching biometric systems through different categories can provide lawmakers, legal scholars, and the general public with a broader understanding of its impacts and contribute to drafting legislation that adequately focuses on some common minimum standards and potential issues each category may present. This can enhance specific rules for developing, deploying, using, and supervising biometric systems according to their impact on privacy and personal data.

This work aims the development of a taxonomy of biometric systems that can clarify and facilitate their understanding and be incorporated into new legislation to regulate better biometric systems' development, deployment, assessment, and accountability. It intends to be a dynamic and technologically neutral tool, capable of avoiding the risks of under- and overinclusion that may result from rapid technological advances.

While the scientific literature on the technical aspects of biometric technologies is extensive and dates back several years, research on the legal aspects of these technologies is comparatively recent and limited. However, it has gained considerable momentum, primarily due to the impact these technologies have on individuals' privacy and personal data protection \parencite{Jain_Kumar_2012}. Thus, this study contributes to the scientific literature on the legal aspects of biometric technologies.

The following section briefly describes biometric systems and some reasons to classify them into smaller groups. In Section \ref{sec:taxonomy}, the proposed taxonomy is presented and discussed. Section \ref{sec:examples} analyses some specific legislations adopted to regulate biometric technologies and how embracing the proposed taxonomy can improve the legal framework. Moreover, in Section \ref{sec:conclusion}, the conclusions from this paper are presented.

\section{Background}
\label{sec:background}

Biometric technologies have long been used in cases where personal identity plays an essential role, and for that reason, numerous applications can benefit from the use of biometric data. At the same time, many legal questions remain to be addressed concerning their deployment.

Many authors have presented some definitions for the term ‘biometric systems’, considering slight differences among them. From some definitions presented in the literature, it is possible to derive a more detailed one: biometric systems are machine-based systems that process biometric data and compare it to a database to (i) identify or (ii) verify the identity or a claim of persons. This comparison can be performed through a recognition process (i) fully automated or (ii) assisted by a human being, based on their distinguishing and repeatable biological (physical and physiological) and behavioural characteristics\footnote{According to the ISO/IEC 2382-37:2022, ``Behavioural and biological characteristics cannot be completely separated which is why the definition uses ‘and’ instead of ‘and/or’.'' }.

In general terms, a biometric system uses a capture device (which can be any piece of hardware and supporting software and firmware and may comprise several components) to collect digital representations (also known as \textit{biometric samples}) of biometric characteristics, to which are then applied an algorithm to convert them into a biometric template. These biometric templates are stored in a database that is accessed when on the following occasions, a biometric sample is presented to the system for comparison. After converting the second biometric sample into a template, a comparison can be executed \parencite{van_der_Ploeg_1999}.

In the case of biometric systems, a taxonomy can provide a deeper understanding of its components, allowing the enactment of legislation focusing on specific aspects according to different interests at the moment (e.g., focusing on specific applications, impact on individual privacy or data protection), instead of exceedingly broad regulations.

The use of taxonomies can contribute to making law's complexities more manageable as it reflects the legal culture of a given legal system, and it can evolve to accommodate doctrinal, legal, and social changes within itself while allowing transferring knowledge from one area to another \parencite{Mattei_1997}. As stated by Birks, it is impossible to achieve legal certainty if and so long as taxonomy is neglected because this neglection leads to errors and confusion \parencite{Birks_1996}.

In the EU, taxonomies have been used, for example, to create a common classification system for sustainable economic activities (i.e., the 'EU Taxonomy') as part of the Union's sustainable finance framework.

The proposed taxonomy is based on extant scientific literature and previous biometric systems classifications. Extant biometric classifications usually are referred to in works focusing on other aspects (generally technical ones) or specific applications of biometric systems and do not delve into the details of the classifications (e.g., \parencite{Singh_Singh_2013,Shyam_Singh_2014,Ferrag_Maglaras_Derhab_Korba_2018}).

It also fills the gap identified on the scientific literature and provides a systematic classification of biometric systems according to their several components, applications, and configurations. One aspect of this problem, related to the absence of systematisation in the nomenclature used, has long been discussed in the literature \parencite{Jain_Ross_2008}.

The decision to develop a taxonomy of biometric systems is based on the fact that taxonomies play an essential role in research and management because the classification of objects helps researchers and practitioners understand and analyse complex domains. It can also order otherwise disorderly concepts and allow researchers to postulate the relationships among the concepts \parencite{Nickerson_Varshney_Muntermann_2013}.

Considering the existence of various options for visualising a taxonomy, e.g., morphological box, hierarchy, or mathematical set \parencite{Szopinski_Schoormann_Kundisch_2020}, the proposed taxonomy is presented using a hierarchical visualisation. since it is well suited for the purpose of structuring and organising knowledge and increasing understanding in the discussion, pedagogy, and research \parencite{Glass_Vessey_1995}.

For practitioners, policymakers, and other stakeholders, the taxonomy gives an overview of different aspects of biometric systems and contributes to the development of legal regulations that can focus on specific aspects, improving the protection of the privacy and personal data of individuals.

\section{Proposed Taxonomy}
\label{sec:taxonomy}

The proposed taxonomy aims to classify biometric systems according to their several components, applications, and configurations. It does not intend to exhaust the topic, nor be a static tool, but rather be a dynamic instrument to contribute to developing better regulations of biometric systems. Other categories shall be included in the future.

\begin{figure}[t]
	\centering
	{\includegraphics[width=9cm]{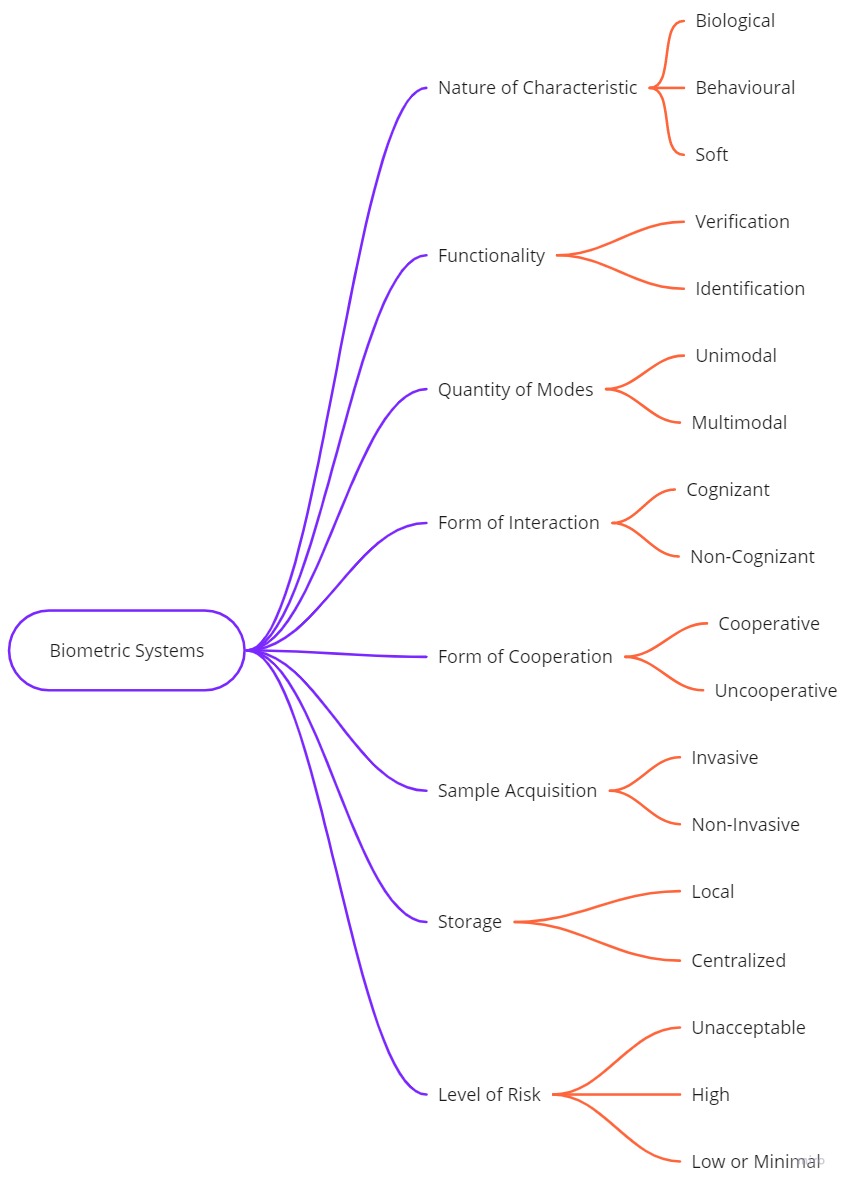}} 
	\caption{Taxonomy of biometric systems.}
	\label{fig:taxonomy} 
\end{figure}

The structure of the proposed taxonomy is presented in Figure \ref{fig:taxonomy}.

\subsection{Nature of Characteristics}
\label{nature}

This category classifies the individual characteristics a biometric system uses to extract repeatable biometric features and recognise a person according to their prevailing nature.

It is often common to describe biometric systems as using 'biological' \parencite{Dargan_Kumar_2020}, 'physical' \parencite{Woodward_1997}, 'physiological' \parencite{Lumini_Nanni_2017,Shaheed_Mao_Qureshi_Kumar_Abbas_Ullah_Zhang_2021}, 'behavioural' \parencite{Jain_Kumar_2012,Ferrag_Maglaras_Derhab_Korba_2018}, 'psychological' \parencite{Schatten_Baca_Rabuzin_2008} and 'psychophysiological' \parencite{Ross_Banerjee_Chen_Chowdhury_Mirjalili_Sharma_Swearingen_Yadav_2019} characteristics.

Although some of these terms may be treated as synonyms at first glance, to develop a taxonomy of biometric systems, it is essential to identify the differences among them. Physical characteristics relate to the body itself, while physiological refer to its functions. Albeit the discussion concerning the separation between physical and biological categories is old \parencite{Haldane_Thompson_Mitchell_Hobhouse_1917}, we recognise the complexities involved and will follow the vocabulary proposed by the ISO/IEC 2382-37:2022, where ‘biological’ encompasses both ‘physical’ and ‘physiological’ characteristics.

Regarding the differences between ‘behavioural’ and ‘psychological’ characteristics, notice that behaviours can be actions (or inactions) or mannerisms, innate or learned, concerning the surrounding environment, meaning a response that could be from internal or external stimuli which bring out the voluntary or involuntary behaviour. On the other hand, the term ‘psychological’ refers to any individual or organism's cognitive and emotional aspects, consisting of how the organism thinks and feels. It ultimately affects the behavioural response. As it is possible to recognise behaviour patterns, and, in essence, a biometric system is a system developed to recognise patterns \parencite{van_den_Broek_2010}, the term ‘behavioural’ is the more suitable to define the nature of a biometric characteristic.

It is also possible to describe another class of characteristics, defined in the literature as ‘soft’ \parencite{Dantcheva_Elia_Ross_2016,Lumini_Nanni_2017} or ‘light’ \parencite{Ailisto_Lindholm_Makela_Vildjiounaite_2004} biometrics. This class encompasses, e.g., age, gender, hair colour, and weight. Although these characteristics alone do not allow the unique identification of a person, they are relevant for the discussion because, according to \textcite{Bisztray_Gruschka_Bourlai_Fritsch_2021}, soft biometrics are not covered by the protection granted by Article 9 of the EU GDPR.

\subsection{Functionality}
\label{functionality}

The biometric recognition process encompasses two main functions: biometric verification and biometric identification \parencite{Prabhakar_Pankanti_Jain_2003}. 

Biometric verification is the process of confirming a claim of identity made by a user of a biometric system by comparing a biometric sample provided by that user with a reference already stored in the system. This mode of operation is also referred to as a ‘one-to-one comparison’ (1:1 comparison), as it compares the acquired information only with those templates corresponding to the claimed identity \parencite{Kindt_2012}, returning a positive result if both information belongs to the same person. For that reason, this function is also called positive. According to \textcite{Dargan_Kumar_2020}, this mode of operation is less expensive and more robust in terms of computation, searching and complexity than identification. It aims to prevent multiple people from using the same identity \parencite{Prabhakar_Pankanti_Jain_2003}.

Biometric identification is the process of comparing a biometric sample submitted to the system with the complete database of biometric references pertaining to all individuals already recorded to retrieve the biometric reference attributable to a single individual. This mode of operation is also referred to as a ‘one-to-many comparison’ (1:N comparison) \parencite{Kindt_2012}. This process is intricated, as it operates by searching for one single user from a database with multiple identities \parencite{Dargan_Kumar_2020}.

According to \textcite{Kindt_2012}, the identification functionality allows checking if someone is enrolled in a particular list or database, which can be a so-called ‘watch list’ or a ‘block list’. However, it does not necessarily provide identity information, only the confirmation that the person is or is not on the list. The output of this process is a list of candidates that can present one or more individuals if the stored characteristic matches with the one presented to the system. It aims to prevent a single person from using multiple identities \parencite{Prabhakar_Pankanti_Jain_2003}.

The biometric verification function presents a lower risk to the privacy of individuals, as it does not require the maintenance of a database of biometric characteristics but only the storage of one specific set of characteristics, which can be done centrally or locally.

\subsection{Quantity of Modes}
\label{modes}

This category classifies the biometric systems according to the number of modes used during the recognition process. According to the ISO/IEC 2382-37:2022, a mode is a combination of a biometric characteristic type, a sensor type, and a processing method. In this context, a biometric system can be unimodal or multimodal.

As the name indicates, a unimodal biometric system encompasses only one biometric characteristic type, sensor type, and processing method \parencite{Roy_Marcel_2010}. This configuration can present several limitations in terms of accuracy, universality, distinctiveness, and acceptability \parencite{Lumini_Nanni_2017}.

On the other hand, a multimodal biometric system is one that combines several biometric modes (a process also known as biometric fusion), which can be a combination of different biometric characteristics types, the use of different sensors for capturing the same biometric characteristic, or the combination of different processing methods. This category of biometric systems can overcome some of the limitations presented by unimodal systems \parencite{Delac_Grgic_2004}.

Considering the protection of personal rights, combining several biometric characteristics or several instances of the same characteristic contributes to higher levels of privacy \parencite{Merkle_Kevenaar_Korte_2012,Anand_Donida_Labati_Genovese_Munoz_Piuri_Scotti_Sforza_2016}. However, it is also argued that multimodal biometric systems present higher security risks as they deal with multiple traits of the same subject \parencite{Lumini_Nanni_2017}.

There are references in the literature on ‘cross-modal’ biometrics, which can be defined as ‘the association of data pertaining to one biometric modality with that of another modality’, aiming to improve the performance of biometric systems \parencite{Nagrani_Albanie_Zisserman_2018,Ross_Banerjee_Chen_Chowdhury_Mirjalili_Sharma_Swearingen_Yadav_2019}.

According to \parencite{Roy_Marcel_2010}, in the case of multimodal biometric systems, the modes are considered separately, while in cross-modal systems, there is the exploitation of information which might be embedded in both modes used by the system. In that sense, it can be understood that a ‘cross-modal’ biometric system is a species of a multimodal biometric system. Simply put, all cross-modal biometric systems are multimodal, but not all multimodal biometric systems are cross-modal.

\subsection{Form of Interaction}
\label{interaction}

An individual's interaction with a biometric system to have their biometric characteristics captured by the system is called a ‘biometric presentation’. This process can be performed with or without their awareness.

According to the ISO/IEC 2382-37:2022 Information technology - Vocabulary - Part 37: Biometrics, the presentation made with the subject's awareness is called a ‘cognizant presentation’, while the presentation made without their awareness is called a ‘non-cognizant presentation’. The ‘non-cognizant presentation’ is often referred to in the literature as being performed ‘without the knowledge of the user’, \parencite{Yampolskiy_Govindaraju_2008}, 'at a distance' \parencite{Choi_2022}, or 'remote biometric' \parencite{Donohue_2012}.

As stated by Article 9 of the EU GDPR, the processing of special categories of personal data, which include biometric data, shall be prohibited, except if performed according to some very strict possibilities, the first of them is the data subject has given \textit{explicit consent} to the processing of their personal data.

This \textit{explicit consent} can only be given if the data subject is aware that a biometric system is processing their personal data.

The permissions for processing given by Article 9(2)(e) and (g), related to ‘personal data which are manifestly made public by the data subject’ and when ‘necessary for reasons of substantial public interest’ are usually invoked to justify the use of remote biometric recognition systems; however, these possibilities are still the focus of intense debate \parencite{Koptelov_2021, Vilanova_Jou_2021}.

\subsection{Form of Cooperation}
\label{cooperation}

Besides the cognisance of the subject discussed in Section \ref{interaction}, the 'biometric presentation' can be carried out with or without the cooperation of the individual.

When the individual is motivated to achieve a successful completion of the acquisition process, which encompasses a series of actions (e.g., to obtain an International Civil Aviation Organisation (ICAO) compliant passport photograph, the individual will have to undertake several steps, e.g. remove glasses, look directly at the camera and not smiling, etc., and then the collected information shall be processed to produce a biometric sample) undertaken to effect a biometric capture, they are called a ‘cooperative subject’.

The cooperative subject may be classified as subversive or non-subversive according to their willing attempt to subvert the correct and intended biometric system policy and avoid being matched to their own biometric reference.

On the other hand, when the individual is motivated not to achieve a successful completion of the acquisition process, they are called ‘uncooperative subjects’. To be an uncooperative subject, they must first be aware that their biometric data is being collected and not provide explicit consent.

In the literature, it is possible to find another classification, known as 'stand-off' biometrics \parencite{Wheeler_Perera_Abramovich_Yu_Tu_2008,Gorodnichy_2009}. This category would include systems capable of operating at a greater‐than‐normal distance between subject and sensor and with less‐constrained subject behaviour \parencite{International_Biometrics_Group_2011}, therefore collecting biometric data with minimal or no direct engagement of the subject and, in many cases, even without their knowledge of the capture process, which can be considered a ‘clandestine use’ of the biometric system \parencite{Woodward_Orlans_Higgins_2003}. This situation would involve a ‘non-cognizant presentation’ but cannot be classified according to the form of cooperation, as to be cooperative or uncooperative, the subject must be aware of the collection process.

\subsection{Sample Acquisition}
\label{acquisition}

The process of obtaining and recording, in a retrievable form, signals of biometric characteristics directly from individuals or from representations of those biometric characteristics and later performing additional processing to attempt to produce a suitable biometric sample is denominated as a ‘biometric acquisition process’.

This biometric acquisition process needs a device to collect the signal from the biometric characteristic and convert it to a captured biometric sample. This device can be any piece of hardware (and supporting software and firmware) and may comprise components such as an illumination source, one or more sensors, etc.

When an individual must interact directly with the device to perform the biometric acquisition process, the biometric system is defined as being ‘invasive’ \parencite{Jain_Kumar_2012} or 'obtrusive' \parencite{Ailisto_Lindholm_Makela_Vildjiounaite_2004,Yampolskiy_Govindaraju_2008}. On the other hand, when this process can be performed without the direct interaction of the individual with the device, it is named ‘non-invasive’ or ‘unobtrusive’.

The obtrusiveness of the biometric acquisition process can interfere with the system’s accuracy by potentially influencing the behaviour of the individuals interacting with the system. Unobtrusive biometric systems have been used to mitigate this issue, especially to assess emotional responses from individuals \parencite{Gonzalez_Viejo_Fuentes_Howell_Torrico_Dunshea_2019,Fuentes_Wong_Gonzalez_Viejo_2020}.

\subsection{Storage}
\label{storage}

The biometric recognition process can be performed according to two main functions, as discussed in Section \ref{functionality}. Typically, this process involves the comparison of incoming biometric samples with records stored as biometric references in a database.

However, biometric systems can depend on several different databases according to the stage of the processing being performed (a general description of these stages can be found in \parencite{Wayman_1997}). For example, a biometric application database stores biometric data (which need not be attributable to a specific subject) and associated metadata developed from and supporting the operation of a biometric application. A biometric enrolment database stores biometric enrolment data records that are attributed to a specific subject and contains non-biometric data associated with biometric reference identifiers. This database can optionally include the biometric reference database, including indexed data records containing biometric references. The merging or unbundling of these databases can be defined by security, privacy, legislation, architecture, performance, or other reasons. 

Depending on the purpose of the biometric system, the templates used for performing the comparisons can be stored in a central database or recorded locally, e.g., on a smart card issued to the individual \parencite{Prabhakar_Pankanti_Jain_2003}. The decision to store the biometric templates in a centralised or localised database will imply different consequences to biometric systems and the protection of collected personal data.

\subsection{Level of Risk}
\label{risk}

In April 2021, the European Commission published a proposal for a regulation laying down harmonised rules on Artificial Intelligence, known as the Artificial Intelligence Act (AI Act). Although not focusing specifically on biometric technologies, this proposal presents several references to biometric systems and a three-tier classification of artificial intelligence practices based on the level of risk to fundamental rights that we considered as a possible categorisation of biometric systems in our taxonomy.

As proposed in the AI Act, the three levels of risk are: (i) unacceptable, (ii) high, and (iii) low or minimal.

AI systems whose use can create unacceptable risks of violating fundamental rights are prohibited from being developed, placed on the market, or used in the EU. An example of a biometric system that fits in this category is the use of ‘real-time’ remote biometric identification systems in publicly accessible spaces for the purpose of law enforcement. Their use is prohibited unless certain limited exceptions apply, as stated in Article 5(1)(d), (2), and (3) of the AI Act.

The AI Act will authorise the placing on the EU market of AI systems considered to be of high risk only if they comply with certain mandatory requirements, as described in Chapter 2 of Title III of the proposal. In this category can be included systems used for biometric identification and categorisation of natural persons, like AI systems intended to be used for the ‘real-time’ and ‘post’ remote biometric identification of natural persons.

Biometric systems whose use is not considered to present an unacceptable or high risk are allowed to be freely used and would be subject to minimum transparency obligations.

\section{Examples of Applications}
\label{sec:examples}

To demonstrate the applicability of the proposed taxonomy, we present some situations where it would be possible to develop regulations focusing on specific aspects of biometric systems, thus providing different legal frameworks according to the system's potential impact on individuals' privacy.

As previously stated, the enactment of broad and generic legislation aiming to protect personal data and individual privacy can result in the banishment or imposition of moratoria or strict requirements on the use of some specific biometric technologies.

To address the issue of government surveillance in the USA, in 2019, the San Francisco Board of Supervisors approved an ordinance banning the ‘acquisition, retention and use of surveillance technology [and] allowing the acquisition and retention of face recognition technology under certain conditions’, making San Francisco the first major city in the United States to ban government use of facial recognition surveillance systems.

Following this example, in 2020, the Boston City Council approved an ordinance banning the use of facial recognition technology by Boston police and other city departments amid evidence that the existing systems misidentify people of colour at an exorbitantly high rate.

More recently, the Baltimore City Council has enacted the City of Baltimore Ordinance 21-038,  which prohibits the ‘Baltimore City government from purchasing or obtaining certain face surveillance technology; […] contracting or subcontracting with another for the purpose of face surveillance technology; prohibiting any person in Baltimore City from obtaining, retaining, accessing, or using certain face surveillance technology or any information obtained from certain face surveillance technology’.

Currently, at least 17 communities across the USA have adopted some kind of local ban on the use of facial recognition systems \parencite{Sheard_Schwartz_2022}. Also in the USA, a coalition of civil society organisations is proposing a federal ban, instead of any form of regulation, on the use of facial recognition technologies by US law enforcement agencies. It resulted, in 2020, in the introduction of a proposal entitled ‘Stop Biometric Surveillance by Law Enforcement Act’ (H.R. 7235), aiming to prohibit the ‘use of facial recognition technology on any image acquired by body-worn cameras of law enforcement officers’.

In Italy, the Data Protection Agency prohibited the use of facial recognition technologies by government agencies until a specific law regulating the issue is adopted, unless the processing is carried out for investigations by the judiciary or the prevention and repression of crimes \parencite{Garante_2022}.

While the use of surveillance technology, particularly facial recognition surveillance systems, can violate individual privacy and inhibit freedom of expression, enacting legislation prohibiting the acquisition and use of a specific biometric technology does not solve the problem, as even more intrusive methods of surveillance are constantly being developed \parencite{Thomas_2019}.

The strictest regulation of facial recognition technologies may cause a switch to one or several of the other forms of remote surveillance technologies currently being developed (e.g., gait analysis or heartbeat signature), which can be even more invasive and harmful to the privacy and protection of personal data and would not be cover by the current prohibitions. On the other hand, attempting to regulate biometric surveillance technologies one by one is likely to be worthless, as the legislative process cannot follow the pace of development of these technologies.

This requires drafting more detailed legislation focusing on the peculiarities of biometric systems to provide a legal framework that adequately protects personal data and privacy without prejudicing the development of new technologies.

The AI Act adopts definitions capable of regulating several biometric systems simultaneously, according to their peculiarities. When prohibiting the use of ‘real time’ remote biometric identification systems in publicly accessible spaces for the purpose of law enforcement, the AI Act does not focus specifically on facial recognition systems but forbids the development, placing on the market, or use in the EU of any biometric system presenting a ‘non-cognizant presentation’, as discussed in Section \ref{interaction}.

Systems developed for biometric identification, such as surveillance systems, pose a greater risk to individual privacy as they require the maintenance of extensive biometric characteristic databases of enrolled individuals. Therefore, they must be subject to more detailed regulation than systems developed only for biometric verification, as discussed in Section \ref{functionality}. Additionally, the biometric identification process does not confirm a specific identity but instead provides a list of candidates whose biometric characteristics resemble the one presented to the system, increasing the risk of false identification, which can be particularly severe in certain circumstances.

Depending on the location of the biometric templates database, it is possible to develop specific rules as well. Systems that maintain smaller, decentralised databases could benefit from softer rules than systems that depend on a bigger, more complex, and centralised database of biometric templates. As the identification process requires the existence of a large, centralised database, perhaps it should observe some additional requirements, such as the need to demonstrate the observation of best cybersecurity practices and the maintenance of a user login ledger. Also, it can be imposed that databases containing biometric images must be encrypted, thus inhibiting their compromise in bulk.

\section{Conclusion}
\label{sec:conclusion}

Despite the increasing adoption of biometric technologies, the related regulation has progressed at a different pace, particularly in safeguarding individuals' privacy and personal data. The widespread deployment of biometric systems has led to various privacy concerns, including unintended functional scope, unintended application scope, and covert surveillance. Nevertheless, these concerns can be mitigated by enacting more suitable legislation.

The implementation of regulations such as the EU GDPR has emphasised the importance of 'designing privacy-preserving methods in the context of biometric systems' \parencite{Ross_Banerjee_Chen_Chowdhury_Mirjalili_Sharma_Swearingen_Yadav_2019}. However, it is equally crucial to establish more comprehensive legislation that specifically targets the unique aspects of these systems, providing a legal framework that effectively safeguards personal data and privacy while still fostering technological advancements.

To address this issue, the proposed taxonomy aims to bridge the gap and aid in the enactment of new regulations to govern the development, deployment, assessment, and accountability of biometric systems. The taxonomy categorises biometric systems based on their various components, applications, and configurations, with the intention of being a dynamic tool that can evolve and expand to serve its purpose continually. Although not exhaustive, the taxonomy will continue to incorporate additional categories as it is evaluated for its effectiveness in promoting better regulation of biometric systems \parencite{Mattei_1997}.

Some of the proposed categories may present sub-divisions, such as the classification presented by \textcite{Yampolskiy_Govindaraju_2008}, where behavioural biometrics are classified into five categories. However, at the moment, we consider further studies necessary to decide on a more elaborated taxonomy.

\printbibliography

\end{document}